\documentclass[reprint,aps,pra,amsmath,superscriptaddress,amssymb,showpacs,footinbib,longbibliography,10pt]{revtex4-1}

\usepackage{graphicx}
\usepackage{hyperref}
\usepackage{braket}
\usepackage{color}

\newcommand{\kv}{\mathbf k}
\newcommand{\eps}{\varepsilon}

\graphicspath{{../}}

\begin{document}

\title{Stripe order and magnetic anisotropy in the $S=1$ antiferromagnet BaMoP$_2$O$_8$}

\author{Jan Hembacher}
\affiliation{Experimental Physics VI, Center for Electronic Correlations and Magnetism, Institute of Physics, University of Augsburg, 86135 Augsburg, Germany}

\author{Danis I. Badrtdinov}
\affiliation{Experimental Physics VI, Center for Electronic Correlations and Magnetism, Institute of Physics, University of Augsburg, 86135 Augsburg, Germany}
\affiliation{Theoretical Physics and Applied Mathematics Department, Ural Federal University, 620002 Yekaterinburg, Russia}

\author{Lei Ding}
\affiliation{Experimental Physics VI, Center for Electronic Correlations and Magnetism, Institute of Physics, University of Augsburg, 86135 Augsburg, Germany}

\author{Zuzanna Sobczak}
\affiliation{Experimental Physics VI, Center for Electronic Correlations and Magnetism, Institute of Physics, University of Augsburg, 86135 Augsburg, Germany}
\affiliation{Faculty of Applied Physics and Mathematics, Gdansk University of Technology, 80-233 Gdansk, Poland}

\author{Clemens Ritter}
\affiliation{Institut Laue-Langevin, BP 156, F-38042 Grenoble, France}

\author{Vladimir V. Mazurenko}
\affiliation{Theoretical Physics and Applied Mathematics Department, Ural Federal University, 620002 Yekaterinburg, Russia}

\author{Alexander A. Tsirlin}
\email{altsirlin@gmail.com}
\affiliation{Experimental Physics VI, Center for Electronic Correlations and Magnetism, Institute of Physics, University of Augsburg, 86135 Augsburg, Germany}
\affiliation{Theoretical Physics and Applied Mathematics Department, Ural Federal University, 620002 Yekaterinburg, Russia}

\begin{abstract}
Magnetic behavior of yavapaiite-type BaMoP$_2$O$_8$ with the spatially anisotropic triangular arrangement of the $S=1$ Mo$^{4+}$ ions is explored using thermodynamic measurements, neutron diffraction, and density-functional band-structure calculations. A broad maximum in the magnetic susceptibility around 46\,K is followed by the stripe antiferromagnetic order with the propagation vector $\kv=(\frac12,\frac12,\frac12)$ formed below $T_N\simeq 21$\,K. This stripe phase is triggered by a pronounced one-dimensionality of the spin lattice, where one of the in-plane couplings, $J_2\simeq 4.6$\,meV, is much stronger than its $J_1\simeq 0.4$\,meV counterpart, and stabilized by the weak easy-axis anisotropy. The ordered moment of 1.42(9)\,$\mu_B$ at 1.5\,K is significantly lower than the spin-only moment of 2\,$\mu_B$ due to a combined effect of quantum fluctuations and spin-orbit coupling. 
\end{abstract}

\maketitle


\section{Introduction}
\label{sec:introduction}
$4d$ and $5d$ transition metals are largely different from their $3d$ counterparts, especially in terms of magnetism. The proclivity to low-spin states and the sizable spin-orbit coupling render $4d$ and $5d$ spins highly anisotropic, giving rise to unusual frustrated scenarios and non-trivial ground states, including the enticing quantum spin liquid phase of Kitaev magnets~\cite{kitaev2006,trebst2017,winter2017,hermanns2018}. The Kitaev model is defined on the honeycomb lattice with nearest-neighbor exchange interactions, where no geometrical frustration occurs, and long-range magnetic order is solely destabilized by the exchange anisotropy (exchange frustration). Combining geometrical and exchange frustration, e.g., on a triangular lattice, may be another interesting direction. It received a thorough theoretical consideration~\cite{jackeli2015,becker2015,li2015,rousochatzakis2016,kos2017}, but has not been realized experimentally yet. A general problem in this case is that $4d$ and $5d$ ions rarely form triangular configurations, and only a few suitable structure types exist. One of them is the yavapaiite structure with triangular layers of octahedrally coordinated transition-metal ions. These ions are typically trivalent, as in the yavapaiite mineral KFe(SO$_4)_2$ and related compounds that serve as good examples of triangular Heisenberg antiferromagnets~\cite{serrano1998,serrano1999,inami2007}.

Here, we explore the magnetic behavior of BaMoP$_2$O$_8$ that also belongs to the yavapaiite family, but features the tetravalent Mo$^{4+}$ ($4d^2$) cation in the place of Fe$^{3+}$ ($3d^5$). As the $4+$ oxidation state is by far more common for $4d$ and $5d$ metals in oxides, BaMoP$_2$O$_8$ could be a promising testing ground for creating triangular arrangements of heavier transition-metal ions. 

\begin{figure}[!h]
\includegraphics[width=0.42\textwidth]{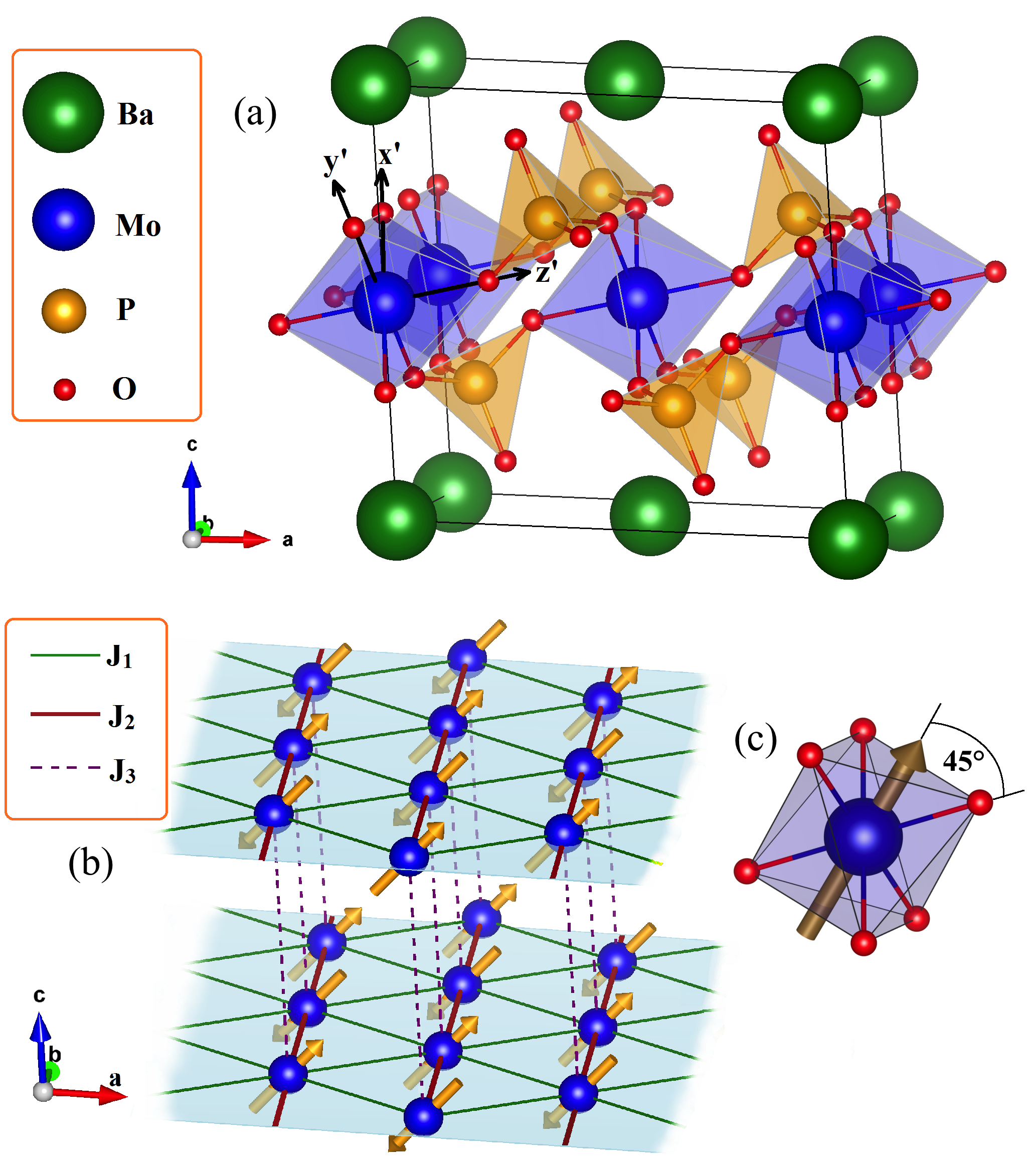}
\caption{(a) Crystal structure of BaMoP$_2$O$_8$. (b) Triangular arrangement of the Mo$^{4+}$ ions and the corresponding magnetic model obtained from \textit{ab initio} calculations. The arrows show stripe antiferromagnetic order determined by neutron diffraction. The $a$ and $b$ vectors of the crystal coordinate frame are along the Cartesian $x$ and $y$ vectors, respectively. The local coordinate frame $x^\prime y^\prime z^\prime$ of the MoO$_6$ octahedron is depicted in the upper panel.  (c) Magnetic moment direction  within the MoO$_6$ octahedron. VESTA software was used for crystal structure visualization~\cite{vesta}.}
\label{fig:Crystal}
\end{figure}
Only the crystal structure of BaMoP$_2$O$_8$ has been reported to date~\cite{raveau1995}. It features layers formed by MoO$_6$ octahedra linked together via PO$_4$ tetrahedra, with Ba atoms separating the layers (Fig.~\ref{fig:Crystal}). The overall structure is remarkably simple and has only one formula unit per cell. Our combined experimental and computational study reveals signatures of quantum magnetism in BaMoP$_2$O$_8$. The broad maximum in the magnetic susceptibility around 46\,K indicates short-range spin order, whereas the N\'eel temperature of $T_N\simeq 21$\,K is  suppressed compared to the Curie-Weiss temperature of about 100\,K. However, these features are accompanied by a sizable distortion of the triangular spin lattice, where the coupling along one direction is predominant. We conclude that BaMoP$_2$O$_8$ can be viewed as a spin-1 chain antiferromagnet, and  we discuss the role of frustration and magnetic anisotropy in this material.

\section{Methods}
\label{sec:methods}

\subsection{Experimental}
Polycrystalline samples of BaMoP$_2$O$_8$ were synthesized using a two-step procedure. First, a mixture of BaCO$_3$, NH$_4$H$_2$PO$_4$, and MoO$_3$ was annealed at 600\,$^{\circ}$C for 24 hours in air. The amount of MoO$_3$ corresponded to two thirds of the stoichiometric amount of Mo in BaMoP$_2$O$_8$. On the second step, the precursor was mixed with metallic molybdenum and annealed in an evacuated and sealed quartz tube at 900\,$^{\circ}$C for 24 hours. Best results were obtained using two-fold excess of metallic molybdenum~\cite{supplement}, which was eventually washed out using 23\,\% HNO$_3$ after the synthesis.

Sample quality was checked by powder x-ray diffraction (XRD). Besides the main phase of BaMoP$_2$O$_8$, only 2.5\,wt.\%  amount of the MoO$_2$ impurity was detected. Magnetic susceptibility of MoO$_2$ is low and weakly temperature-dependent within the temperature range of our study~\cite{ghose1976}, so its contribution can be safely neglected. Lab XRD data were collected on the Rigaku MiniFlex and PANalytical Empyrean diffractometers (Bragg-Brentano geometry, CuK$_{\alpha}$ radiation). Several representative samples were further measured at the ID22 beamline of the European Synchrotron Radiation Facility (ESRF, Grenoble, $\lambda=0.4002$\,\r A), where the capillary geometry with point detectors preceded by Si analyzer crystals was used. The measurements were performed at room temperature and 20\,K using He cryostat. The small reflection width in the synchrotron measurements ($\Delta(2\theta)\simeq 0.01^{\circ}$) confirmed the absence of major structural defects, such as stacking faults. 

Magnetic susceptibility was measured between 1.8\,K and 350\,K using the MPMS SQUID magnetometer from Quantum Design. Heat capacity was measured with Quantum Design PPMS using the relaxation method.

Neutron diffraction data were collected on a 5\,g powder sample at the D20 diffractometer equipped with the Orange cryostat at the ILL, Grenoble ($\lambda=2.42$\,\r A). Additionally, high-resolution room-temperature data were collected on D2B ($\lambda=1.594$\,\r A). Jana2006~\cite{jana2006} and Fullprof~\cite{fullprof} were used for the crystal and magnetic structure refinement, respectively.

\subsection{Computational}
Density-functional (DFT) band-structure calculations were performed within the generalized gradient approximation (GGA)~\cite{pbe96}. Quantum Espresso~\cite{espresso} and Vienna ab initio Simulation Package (VASP)~\cite{vasp1,vasp2} codes were used. In these calculations, we set the energy cutoff in the plane-wave construction to 400\,eV and the energy convergence criteria to 10$^{-6}$\,eV. For the Brillouin-zone integration, a 8$\times$8$\times$8 Monkhorst-Pack mesh was used. 

Correlations effects were taken into account on the mean-field level using DFT+$U$~\cite{anisimov1991}. The difference between the on-site Coulomb repulsion $U$ and Hund's coupling $J_H$ was obtained by the linear-response method~\cite{cococcioni2005} resulting in $U-J_H=2.0-2.5$\,eV. We fixed $J_H=0.8$\,eV and $U=3$\,eV and used these values in all calculations. Our parametrization is similar to the previous DFT+$U$ reports on molybdates~\cite{shinaoka2013,iqbal2017} and other $4d$ oxides~\cite{lee2006}. 

Magnetic behavior of BaMoP$_2$O$_8$ is described by the spin Hamiltonian
\begin{eqnarray}
\mathcal {\hat{H}} = \sum\limits_{i<j}J_{ij}\hat{\mathbf{S}}_i\hat{\mathbf{S}}_j + \sum\limits_{i \le j}\hat{\mathbf{S}}^{\mu}_i \Delta^{\mu \nu}_{ij}\hat{\mathbf{S}}^{\nu}_j + g\mu_{B}\sum\limits_{i}\hat{\mathbf{S}}_i\mathbf{B},
\label{eq:Magnetic_model}
\end{eqnarray} 
where $J_{ij}$ are isotropic exchange interactions between the spins, and $\mathbf{B}$ is the external magnetic field. The matrix $\Delta^{\mu \nu}_{ij}$ stands for the traceless anisotropic intersite term $\Gamma^{\mu \nu}_{ij}$ at $i \neq j$ and for the single-ion term $A^{\mu \nu}_{i}$ at $i = j$. The latter part appears due to the fact that the Mo$^{4+}$ ions have spin $S=1$. The parameters of Eq.~\eqref{eq:Magnetic_model} are defined with respect to the Cartesian coordinate frame $xyz$, where $x$ and $y$ are along $a$ and $b$, respectively, whereas $z$ deviates from $c$ because of the monoclinic symmetry.

Isotropic exchange integrals $J_{ij}$ were calculated from total energies of collinear spin configurations by a mapping procedure~\cite{xiang2011}, utilizing a supercell with eight molybdenium atoms. Alternatively, local force theorem can be used, and the exchange integrals are given by~\cite{liechtenstein1987,mazurenko2005}
\begin{eqnarray*}
J_{ij} = \frac{1}{2 \pi S^2}\times& \\[5pt]
 \times \int \limits_{-\infty}^{E_F} d \epsilon&\,{\rm Im} \left( \sum \limits_{m, m^{\prime},  n, n^{\prime}} \Delta^{m m^{\prime}}_i G^{m^{\prime} n}_{ij \downarrow} (\epsilon) \Delta^{n n^{\prime}}_j G^{n^{\prime} m}_{ji \uparrow} (\epsilon) \right),
\label{eq:Exchange}
\end{eqnarray*} 
where $m, m^{\prime},  n, n^{\prime}$ are orbital quantum numbers, $S$ is the spin quantum number, and $\Delta^{m m^{\prime}}_i = H^{m m^{\prime}}_{ii \uparrow} - H^{m m^{\prime}}_{ii \downarrow}$ is the on-site potential. In turn, the one-particle Green's function is defined as
\begin{eqnarray}
G^{m m^{\prime}}_{ij \sigma}(\epsilon) = \frac{1}{N_K} \sum \limits_{\mathbf{k},l}^M \frac{c^{ml}_{i \sigma}(\mathbf{k}) c^{m^{\prime}l^*}_{j \sigma}(\bf{k})}{\epsilon - E^{l}_{\sigma}(\mathbf{k})}
\label{eq:Green}
\end{eqnarray} 
In this equation, $c^{ml}_{i \sigma}(\mathbf{k})$ stands for the component of the $l^{\rm th}$ eigenstate and $E^{l}_{\sigma}(\mathbf{k})$ is the corresponding eigenvalue. Both quantities are obtained from the electronic structure, where electronic correlations have been taken into account within DFT+$U$, and thus they inherit the underlying Coulomb repulsion. The summation runs within the first Brillouin zone with the total amount of $k$-points $N_K$ and in all states $M$ involved in the calculation. This approach yields not only total values of the exchange couplings, but also their partial, orbital-resolved contributions.

Magnetic susceptibility of the $S=1$ chain was obtained using the stochastic series expansion (SSE)~\cite{sse} method implemented in the \texttt{loop}~\cite{loop} algorithm of the ALPS~\cite{ALPS} simulation package. We performed simulations for chains with the length $L\le 100$ and periodic boundary conditions.

\section{Results}
\subsection{\label{sec:thermodynamic}Thermodynamic properties}
\begin{figure}[!h]
\includegraphics[width=0.49\textwidth]{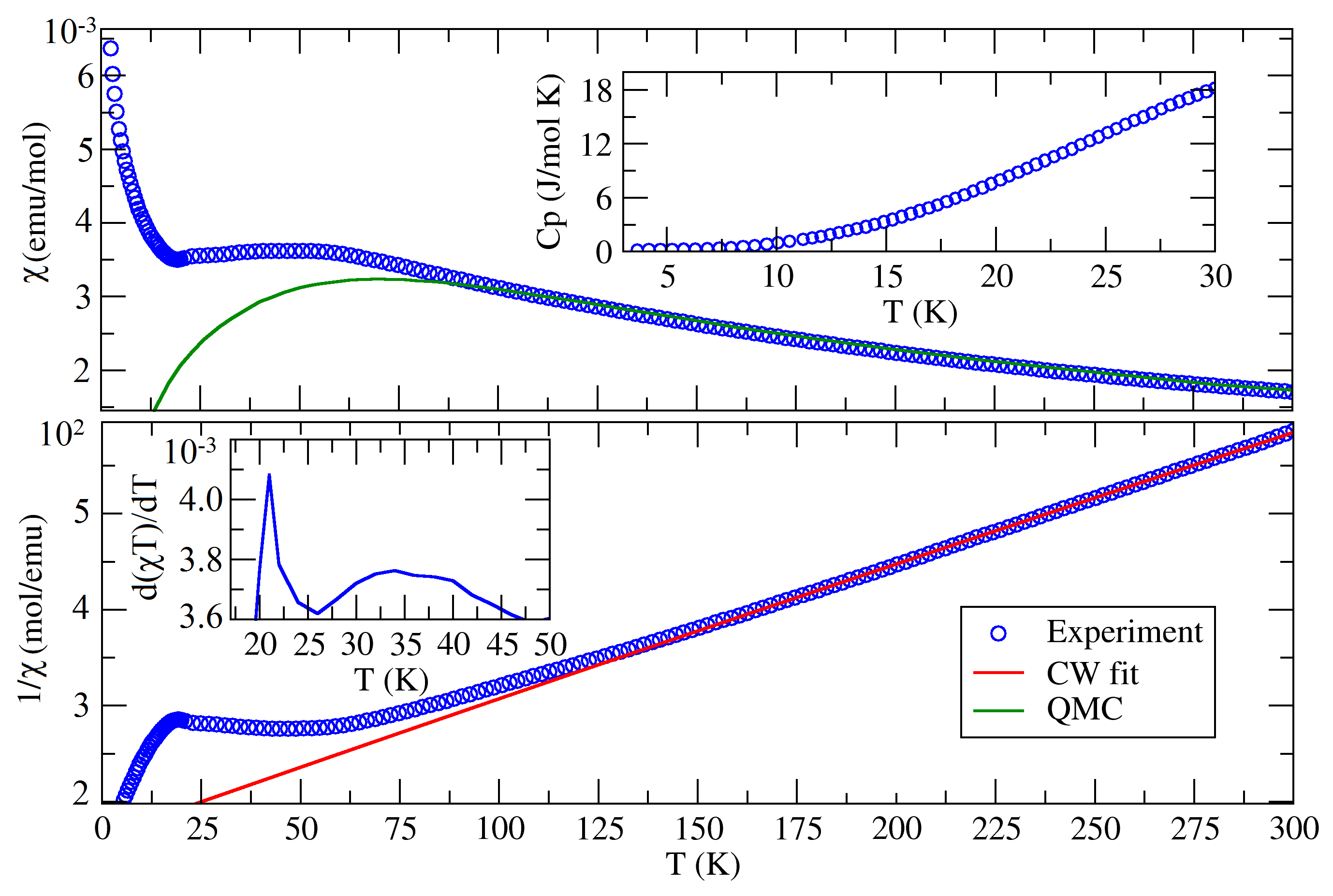}
\caption{ (Top) Magnetic susceptibility of BaMoP$_2$O$_8$. The green line shows the QMC simulation for the $S=1$ chain model. (Bottom) The temperature dependence of $1/\chi(T)$ and corresponding Curie-Weiss fit. The insets show the specific heat $C_p$ and Fisher's heat capacity $d(\chi T)/dT$ as function of temperature. }
\label{fig:Chi}
\end{figure}

Temperature-dependent magnetic susceptibility of BaMoP$_2$O$_8$ is shown in Fig.~\ref{fig:Chi}. A broad maximum around 46\,K is followed by a minimum around 21\,K and a sharp upturn at low temperatures. This upturn shows a strong field dependence typical of a paramagnetic (Curie-like) impurity contribution. The susceptibility curve is rather smooth even around the minimum. However, Fisher's heat capacity $d(\chi T)/dT$ reveals a weak kink at $T_N=21$\,K that, as we confirm below (Sec.~\ref{sec:neutron}), marks the transition into the long-range-ordered state. Interestingly, no transition anomaly could be seen in the zero-field heat-capacity data (Fig.~\ref{fig:Chi}), which are dominated by the lattice contribution and do not show any characteristic signatures of low-dimensional magnetism. 

Inverse susceptibility is linear above 150\,K and identifies the Curie-Weiss regime. The fit with $\chi=\chi_0+C/(T+\Theta)$  in the range $150-300$\,K yields the temperature-independent part $\chi_{0} =0.80 \pm 0.08 \times 10^{-4}$\,emu/mol), Curie-Weiss temperature $\Theta = -111 \pm 11 $\,K, and Curie constant $C=0.67 \pm 0.06 $\,emu\,K/mol. The negative Curie-Weiss temperature reveals predominant antiferromagnetic (AFM) interactions. 

The Curie constant corresponds to the effective moment of 2.31 $\,\mu_B$ to be compared with the spin-only value of 2.83\,$\mu_B$ for the $S=1$ Mo$^{4+}$ ion. This discrepancy implies the presence of an orbital moment that, according to Hund's rules, should be opposite to the spin moment for the less than half-filled shell of the $4d^2$ ion, and thus reduces $\mu_{\rm eff}$. Using $C=N_A(g\mu_B)^2S(S+1)/3k_B$, one arrives at $g=1.63$.

The $\theta/T_N$ ratio of 5.3 suggests that the long-range magnetic order in BaMoP$_2$O$_8$ is strongly impeded. Indeed, the magnetic transition takes place well below the susceptibility maximum. This suppression of the $T_N$ implies that only a small amount of the magnetic entropy is available at the transition, thus rendering the $\lambda$-type anomaly of the specific heat diminutively small. The absence of the transition anomaly in the specific-heat data of quasi-2D antiferromagnets has been demonstrated both experimentally~\cite{lancaster2007,tsirlin2013} and theoretically~\cite{sengupta2003}. In this case, magnetic susceptibility (and the associated Fisher's heat capacity) serves as a more sensitive probe of the transition, because the susceptibility kink is driven by the anisotropy of the ordered state and does not depend on the dimensionality. Microscopic probes are even more useful~\cite{lancaster2007}.

\subsection{\label{sec:neutron}Magnetic structure}
\begin{figure}
\includegraphics[width=0.50\textwidth]{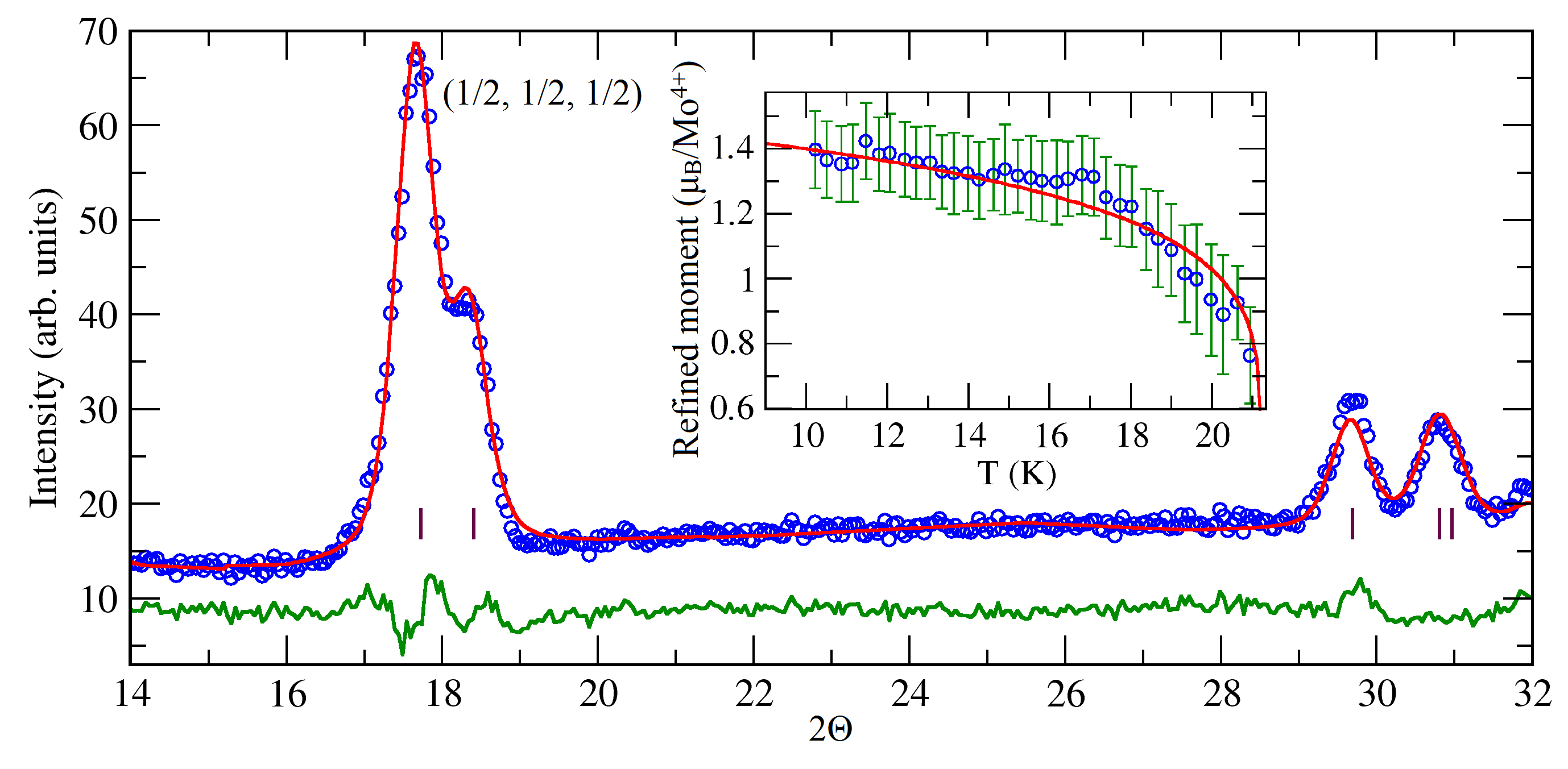}
\caption{Magnetic neutron scattering obtained by subtracting the 25\,K data (above $T_N$) from the 1.5\,K data (below $T_N$). All magnetic peaks are indexed with the propagation vector $(\frac12,\frac12,\frac12)$. The red line is the fit with the covalent form-factor for the magnetic structure shown in Fig.~\ref{fig:Crystal} (bottom), and the green line is the difference. The insert shows temperature dependence of the ordered moment and its empirical fit, as described in the text. 
}
\label{fig:Neutron}
\end{figure}

Neutron diffraction data collected at 1.5\,K and 25\,K look very similar, but the difference pattern (Fig.~\ref{fig:Neutron}) reveals several clear magnetic peaks that could be indexed with the propagation vector $\kv=(\frac12,\frac12,\frac12)$. These peaks gradually weaken upon heating and disappear around 21\,K, thus confirming the formation of long-range magnetic order.

\begin{table}
\caption{\label{tab:neutron}
Magnetic structure refinement using the difference data, $I_{\rm 1.5\,K}-I_{\rm 25\,K}$, and the ionic as well as covalent (\textit{ab initio}) form-factors for Mo$^{4+}$. Magnetic moments $\mu$ and their $x$- and $z$-components are in $\mu_B$, whereas the $\mu_y$ component was zero within the error bar. The normalized vector of the spin direction is $\boldsymbol{\gamma} \sim$ (0.57, 0, 0.82) based on the refinement with the covalent form-factor. $R_{\rm mag}$ is the refinement residual.
}
\begin{ruledtabular}
\begin{tabular}{ccccc}
form-factor & $\mu_x$ & $\mu_z$ & $\mu$  & $R_{\rm mag}$ \\
  \hline
	ionic    & 0.82(3) & 0.84(3) & 1.17(7) & 0.121 \\
	covalent & 0.81(4) & 1.16(3) & 1.42(9) & 0.062 \\
\end{tabular}
\end{ruledtabular}
\end{table}

The refined magnetic structure features antiferromagnetic (AFM) order between the triangular planes. The order within the planes is stripe-type, with stripes of parallel spins along the $[110]$ direction and antiparallel spins along the $[010]$ and $[1\bar 10]$ directions (Fig.~\ref{fig:Crystal}).  Details of the refinement depend on the magnetic form-factor, which has not been reported for Mo$^{4+}$ to date. We thus used the ionic approximation derived for Mo$^{5+}$ in Ref.~\onlinecite{ishikawa2017}. Alternatively, the covalent form-factor is calculated \textit{ab initio}~\cite{FormFactor} and includes effects of both Mo $4d$ and O $2p$ orbitals, as further explained in the next section. Both form-factors lead to very similar magnetic moment directions, but different fit quality and size of the ordered moment (Table~\ref{tab:neutron}). The ionic form-factor yields the higher refinement residual and the lower ordered moment, because all spin density is concentrated on Mo. The covalent form-factor additionally accounts for the spin density on oxygens and should be more realistic.

By monitoring the ordered moment as a function of temperature and using the empirical fit with $\mu=\mu_0(1-T/T_N)^\beta$ we arrive at $\beta = 0.14(1)$
and $T_N = 21.24(3)$\,K. This estimate of the $T_N$ is well in line with the anomaly in Fisher's heat capacity (Fig.~\ref{fig:Chi}). The $\beta$ value should not be confused with the true critical exponent for the magnetization, because we fit the data in the broad temperature range. More sensitive probes like nuclear magnetic resonance or single-crystal neutron diffraction would be needed to extract the true critical behavior in the vicinity of $T_N$.

Magnetic reflections of BaMoP$_2$O$_8$ are broader than the nuclear ones. Given the high crystallinity of our samples, as probed by synchrotron XRD, we can use the width of nuclear reflections as instrumental broadening and estimate the integral breadth associated with the additional size broadening as $\zeta=0.36^{\circ}$ for the strongest magnetic reflection at $2\theta=17.6^{\circ}$. This corresponds to the domain size of 39\,nm according to Scherrer's formula. We thus find that the magnetic order in BaMoP$_2$O$_8$ breaks down into relatively small domains, possibly due to frustration of the underlying spin lattice, as further discussed in Sec.~\ref{sec:discussion}.

\subsection{Electronic structure}
\begin{figure}[!h]
\includegraphics[width=0.47\textwidth]{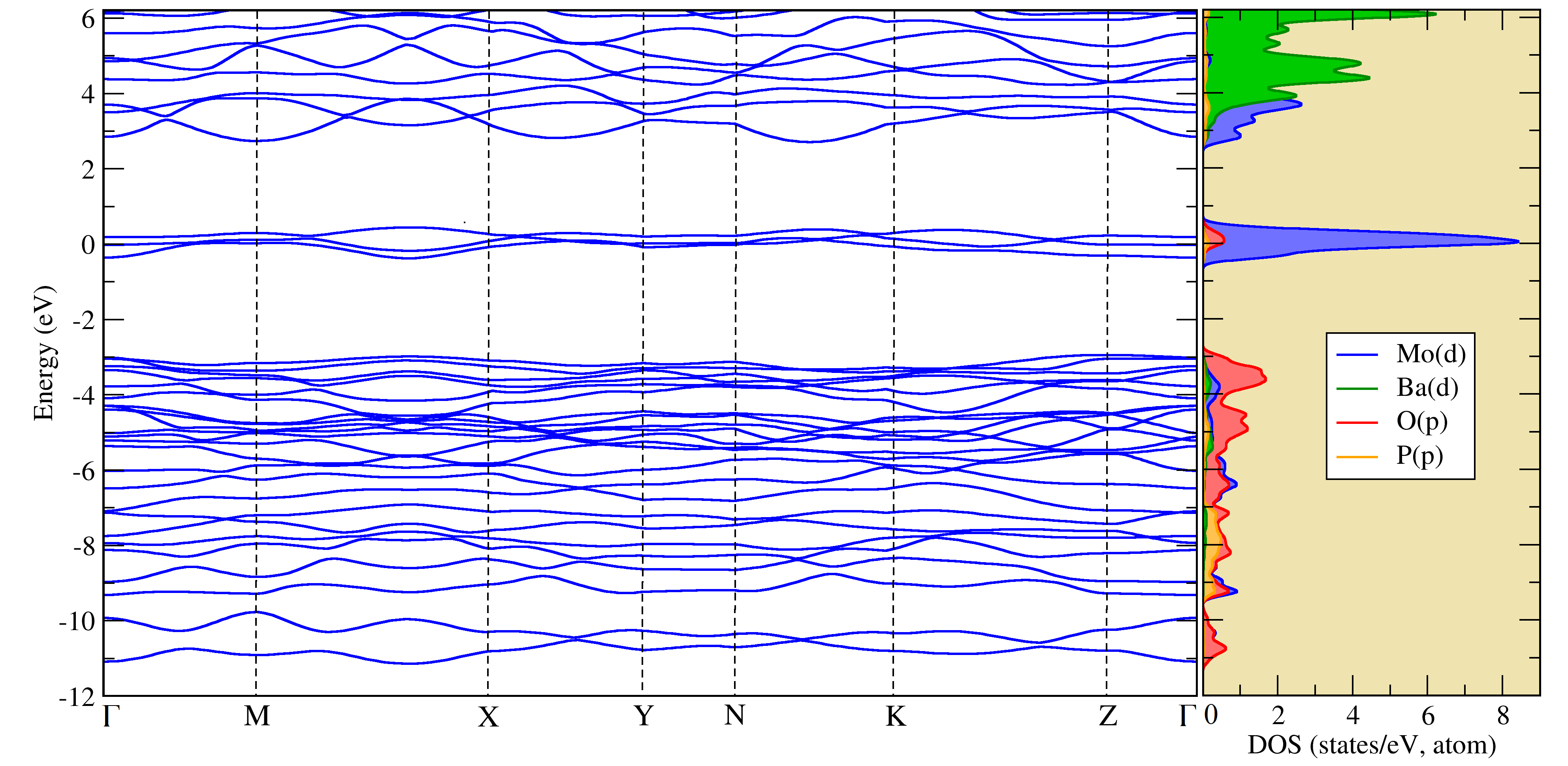}
\caption{ (Left panel) GGA band structure of BaMoP$_2$O$_8$. The high-symmetry $k$-points of the first Brillouin zone are defined as: $\Gamma=(0, 0 ,0)$, $X=(0.5, 0, 0)$, $M=(0.5, 0.778, 0)$, $Y = (0, 0.778, 0)$ , $N = (0, 0.778, 0.525)$, $K = (0.5, 0.778, 0.525)$, $Z = (0, 0, 0.525)$, where all coordinates are given in crystal coordinates in units of $2\pi/a$. (Right panel) Corresponding density of states (DOS) with the atomic contributions. The Fermi level is at zero energy.}
\label{fig:Bands}
\end{figure}

The uncorrelated (GGA) band structure of BaMoP$_2$O$_8$ is shown in Fig.~\ref{fig:Bands}. Its apparent metallicity is due to the fact that Coulomb correlations responsible for opening the band gap in a Mott insulator were not taken into account. The bands below $-3$\,eV mostly comprise O $2p$ states, the bands at the Fermi level are due to the $t_{2g}$ orbitals of Mo, whereas the bands above 2.5\,eV are formed by the $e_g$ orbitals of Mo followed by Ba $5d$.

The $d$-levels are weakly split within the $t_{2g}$ manifold. Using the tight-binding model parametrized via maximally localized Wannier functions~\cite{marzari1997}, we obtain orbital energies of $\eps_{x'y'}$ = 17\,meV, $\eps_{x'z'}$ = 116\,meV, and $\eps_{y'z'}$ = 110\,meV. They are consistent with crystal-field levels expected from the weak local distortion of the MoO$_6$ octahedra that feature two shorter Mo--O distances of 1.95\,\r A and four longer distances of 2.03\,\r A. We define the local coordinate frame $x'y'z'$, where the $z'$ axis is directed along the shorter axial Mo--O bonds and the $x'$ and $y'$ axes are directed along the longer equatorial bonds (Fig.~\ref{fig:Crystal}). Then, the distortion decreases the energy of the $\ket{x'y'}$ orbital lying within the plane formed by the longer Mo--O bonds, and it increases the energies of the $\ket{x'z'}$ and $\ket{y'z'}$ orbitals that do not lie in this plane. Placing two electrons onto such orbitals leads to an orbitally-degenerate scenario, because the $\ket{x'z'}$ and $\ket{y'z'}$ states are very close in energy. Whereas one electron should occupy the $\ket{x'y'}$ state, the second electron can choose between $\ket{x'z'}$ and $\ket{y'z'}$.

The ground-state orbital configuration was obtained from DFT+$U$+SO calculations~\footnote{Similar results are obtained withing DFT+$U$ without including the spin-orbit coupling.}  considering a ferromagnetic spin state by computing the matrix of orbital occupation numbers as
\begin{equation}
 n^{m m^{\prime}} = -\frac{1}{\pi}\, {\rm Im} \left( {\rm } \int \limits_{-\infty}^{E_F} G^{m m^{\prime}}_{ii}(\epsilon) d \epsilon \right). 
\end{equation}
We find that the $\ket{x'y'}$ state hosts one electron indeed, whereas the second electron occupies the mixed $\ket{\varphi_1}=\frac{1}{\sqrt{2}} ( \ket{x'z'} + \ket{y'z'})$ state. Back on the GGA level, such a linear combination of the $\ket{x'z'}$ and $\ket{y'z'}$ has a lower energy of 60\,meV, whereas its orthogonal counterpart, $\ket{\varphi_2}=\frac{1}{\sqrt{2}} ( \ket{x'z'} - \ket{y'z'})$, lies higher at 166\,meV. This splitting between $\ket{\varphi_1}$ and $\ket{\varphi_2}$ can be traced back to the scissor-like distortion in the $x'y'$ plane, where the O--Mo--O angles are $85.2^{\circ}$ and $94.8^{\circ}$, rendering the $x'+y'$ ($\varphi_1$) and $x'-y'$ ($\varphi_2$) directions nonequivalent. However, the lower-energy $\ket{\varphi_1}$ state corresponds to the direction that bisects the angle of $85.2^{\circ}$, where oxygens are closer to the $d$-orbital than in the $\ket{\varphi_2}$ state that follows the $x'-y'$ direction bisecting the angle of $94.8^{\circ}$. More distant neighbors may play a role here, as in the yavapaiite-type KTi(SO$_4)_2$~\cite{nilsen2015}, where the selection of the ground-state orbital also contradicts geometrical crystal-field arguments.

\begin{figure}
\includegraphics[width=0.45\textwidth]{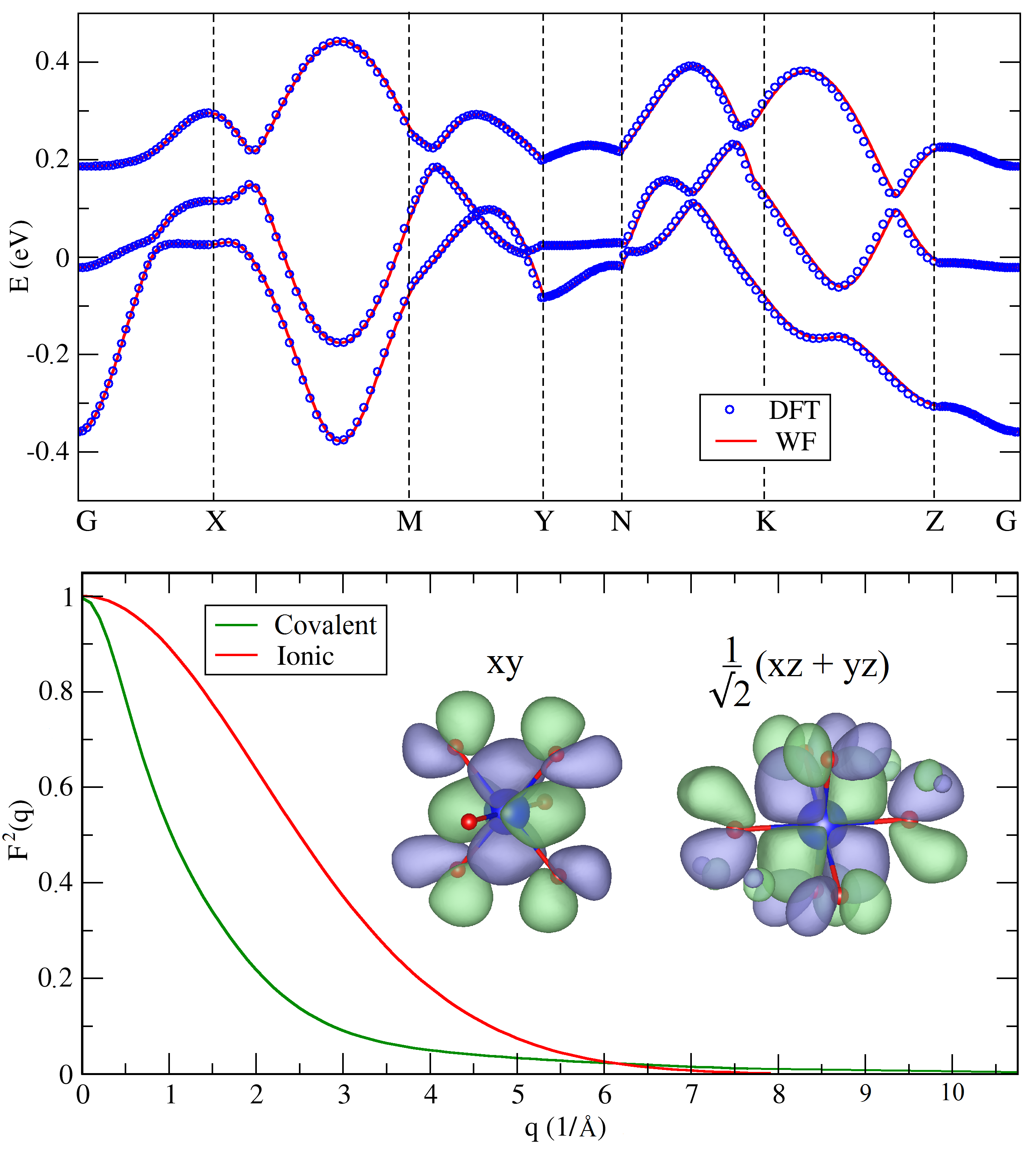}
\caption{ (Top panel) Bands near the Fermi level and their fit using Wannier functions. (Bottom panel) Squared ionic~\cite{ishikawa2017} and covalent magnetic form factors of Mo$^{4+}$. The insert shows the combination of Wannier functions contributing to the magnetic form factor, $\ket{x'y'}$ and $\ket{\varphi_1}=\frac{1}{\sqrt{2}} ( \ket{x'z'} + \ket{y'z'})$.}
\label{fig:WF}
\end{figure}

The Wannier functions associated with the half-filled states $\ket{x'y'}$ and $\ket{\varphi_1}$ are shown in Fig.~\ref{fig:WF}. They are further used to obtain the covalent magnetic form-factor that, in the standard parametrization~\cite{PJBrown}, is given by $A=0.31$, $a=93.14$, $B=0.53$, $b=31.01$, $C=0.07$, $c=292.09$, and $D=0.09$. In Fig.~\ref{fig:WF}, we compare the $q$-dependence of the ionic and covalent form-factors for Mo$^{4+}$. The covalent form-factor is more localized in the reciprocal space, and thus, it has a larger span in the real space, because O $2p$ orbitals admixed to the Mo $4d$ are included.

The DFT+$U$+SO calculations not only lift the orbital degeneracy and choose $\ket{x'y'}$ and $\ket{\varphi_1}$ as the half-filled states, but they also restore the insulating energy spectrum with a band gap of about 2\,eV. The total spin moment within the unit cell containing one Mo atom is equal to 2\,$\mu_B$ according to the $S=1$ nature of Mo$^{4+}$. By placing the spin along the experimental direction $\boldsymbol{\gamma}$ (Table~\ref{tab:neutron}), we find that the spin-orbit coupling generates a sizable orbital moment of 0.35\,$\mu_{B}$ directed opposite to the spin moment. The resulting total magnetic moment $\mu_{\rm DFT}=gS\mu_B=1.65$\,$\mu_B$ corresponds to $g=1.65$ in excellent agreement with $g=1.63$ from the Curie-Weiss fit. On the other hand, $\mu_{\rm DFT}$ is clearly larger than ordered moments obtained from neutron diffraction. Part of this discrepancy is due to the spread of the spin density onto oxygen. However, even the full account of the oxygen atoms via the covalent form-factor leaves $\mu_{\rm DFT}$ and $g=1.63$ from the Curie-Weiss fit somewhat higher than the ordered moment of 1.42(9)\,$\mu_B$ determined experimentally. This remaining discrepancy can be ascribed to quantum fluctuations in the low-dimensional and frustrated spin lattice of BaMoP$_2$O$_8$.

\subsection{Isotropic magnetic interactions}
To determine magnetic interactions in BaMoP$_2$O$_8$, we first analyze electron hoppings within the $t_{2g}$ manifold. In the case of $J_1$, the hoppings between the $\ket{x'y'}$ and $\ket{\varphi_1}$ states are 0.7\,meV and 51.9\,meV, respectively. Much larger hoppings of, respectively, $-153.7$\,meV and 35.4\,meV are found in the case of $J_2$. No significant hoppings beyond nearest neighbors are observed. Therefore, the magnetic model can be restricted to only two in-plane couplings, $J_1$ and $J_2$. Additionally, we consider the out-of-plane coupling $J_3$ that is responsible for the magnetic order along the $c$ direction. Here, the hopping between the $\ket{x'y'}$ states is $-20.5$\,meV, and that between the $\ket{\varphi_1}$ states is negligible. 

\begin{table}[!h]
\centering
\caption [Bset]{Interatomic distances $d_{\rm Mo-Mo}$ (in\,\r A) and isotropic exchange interactions (in\,meV) in BaMoP$_2$O$_8$ calculated using local force theorem (Green's functions technique) for different orbital manifolds ($J^{G}_{\rm I-III}$, see text for details) and by a mapping procedure from total energies ($J^{E}$). The corresponding interaction paths are visualized in Fig.~\ref{fig:Crystal} (bottom). }
\label {basisset}
\begin{ruledtabular}
\begin{tabular}{lcrrrr}
 & $d_{\rm Mo-Mo}$ & $J^{G}_{\rm I}$  & $J^{G}_{\rm II}$  & $J^{G}_{\rm III}$  & $J^{E}$   \\
  \hline
 $J_1$ &  4.880   & 0.4   & 0.3  & 0.2  &  0.4 \\
 $J_2$ &  5.275   & 7.7   & 5.1  & 4.7  &  4.6  \\ 
 $J_3$ &  7.816   & 0.2   & 0.0  & 0.0  &  0.2  \\
\end{tabular}
\end{ruledtabular}
\label{tab:Isotropic_exchange}
\end{table}
Total exchange couplings $J_i$ obtained by a mapping procedure are listed in the last column of Table~\ref{tab:Isotropic_exchange}. As expected from the size of the hopping elements, the in-plane coupling $J_2$ is much larger than $J_1$. The interplane coupling $J_3$ is comparable to $J_1$ and also weak, as expected from the layered nature of the crystal structure. 

\begin{figure}[!h]
\includegraphics[width=0.50\textwidth]{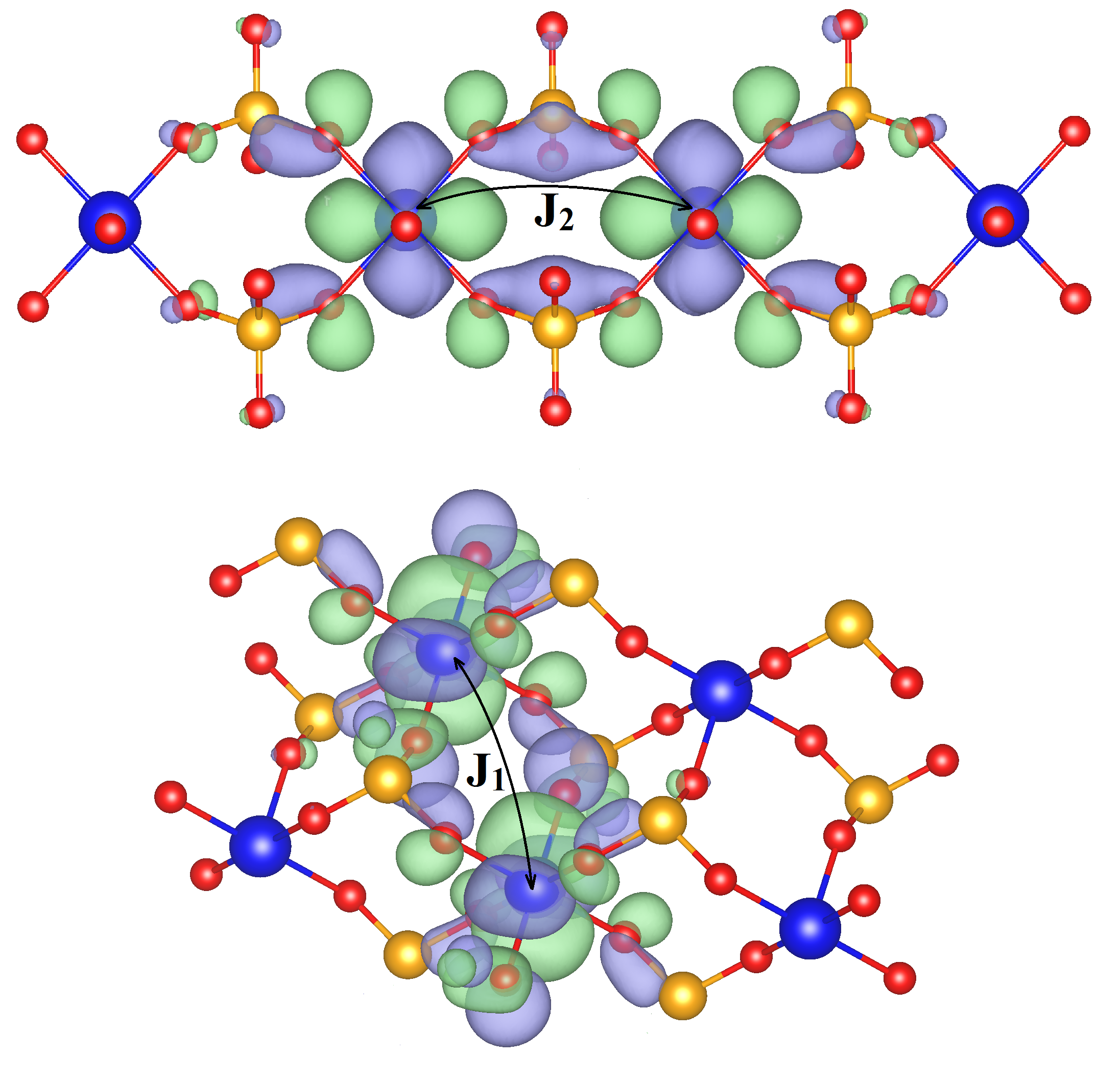}
\caption{ The superexchange pathways for $J_1$ and $J_2$ with Wannier functions based on the Mo $\varphi_1$ and $x'y'$ orbitals, respectively. Different colors indicate different phases of the Wannier functions. }
\label{fig:Overlap}
\end{figure}
To explore the striking difference between $J_2$ and $J_1$, we take advantage of the local force theorem and construct Green's functions from the Wannier functions for different orbital manifolds: (I) only Mo$(t_{2g})$, (II) Mo$(t_{2g})$ + O$(p)$, and (III) Mo$(t_{2g}+e_g)$ + O$(p)$ states. The resulting exchange couplings are also listed in Table~\ref{tab:Isotropic_exchange}. Whereas model I overestimates $J_2$, the combination of Mo $t_{2g}$ and O $p$ states is already sufficient to reproduce the microscopic scenario. A closer examination of different orbital contributions shows that the largest contribution to $J_2$ originates from the $x'y'$ orbitals ($\sim$ 94\%) that overlap on the oxygen atoms of the PO$_4$ tetrahedra (Fig.~\ref{fig:Overlap}), while the interaction $J_1$ is mostly due to the $\varphi_1$ orbitals. 

This interaction mechanism is common to transition-metal phosphates, where the magnitude of the coupling is determined by the linearity of the superexchange pathway~\cite{roca1998,petit2003}. For example, lateral and vertical displacements of the metal-oxygen polyhedra are known to be important in V$^{4+}$ phosphates~\cite{tsirlin2008} and Cu$^{2+}$ phosphates~\cite{nath2014}.
Similar arguments can be applied to our case. The coupling $J_2$ runs along the $b$-direction (two-fold symmetry axis), such that both lateral and vertical displacements are zero. In contrast, the coupling $J_1$ features sizable lateral and vertical displacements allowed by symmetry. Indeed, the overlap of the oxygen "tails" is much stronger in the case of $J_2$, which renders this coupling dominant compared to $J_1$ (Fig.~\ref{fig:Overlap}). 

Our exchange couplings yield the Curie-Weiss temperature $\Theta=-2(4J_1 + 2J_2 + 2J_3)/(3k_B) \sim -88$\,K, in reasonable agreement with the experimental value of $-111$\,K. The $J_2\gg J_1,J_3$ regime leads to the $S=1$ uniform (Haldane) chain as the minimum magnetic model. Indeed, simulated magnetic susceptibility for such a chain with $J_2=4.6$\,meV from Table~\ref{tab:Isotropic_exchange} describes the experimental magnetic susceptibility down to 90\,K with the effective $g$-factor of $g=1.63$ (Fig.~\ref{fig:Chi}). Deviations at lower temperatures may be due to the interchain couplings $J_1$ and $J_3$. 

Weak interchain exchange interactions are responsible for the formation of the long-range magnetic order below $T_N/[J_2 S(S+1)]\simeq 0.2$. According to Ref.~\citep{yasuda2005}, this value of $T_N$ corresponds to $J^{\prime}/J_2 \sim $\,0.1, where $J^{\prime}$ is an effective interchain exchange interaction in the quasi-one-dimensional limit. The ensuing $J^\prime \sim$ 0.46\,meV is of the same order as $J_1$ and $J_3$. Therefore, the interchain couplings obtained \textit{ab initio} are compatible with the formation of the long-range order below $T_N\simeq 21$\,K.

\subsection{Magnetic anisotropy}
\begin{figure}
\includegraphics[width=0.48\textwidth]{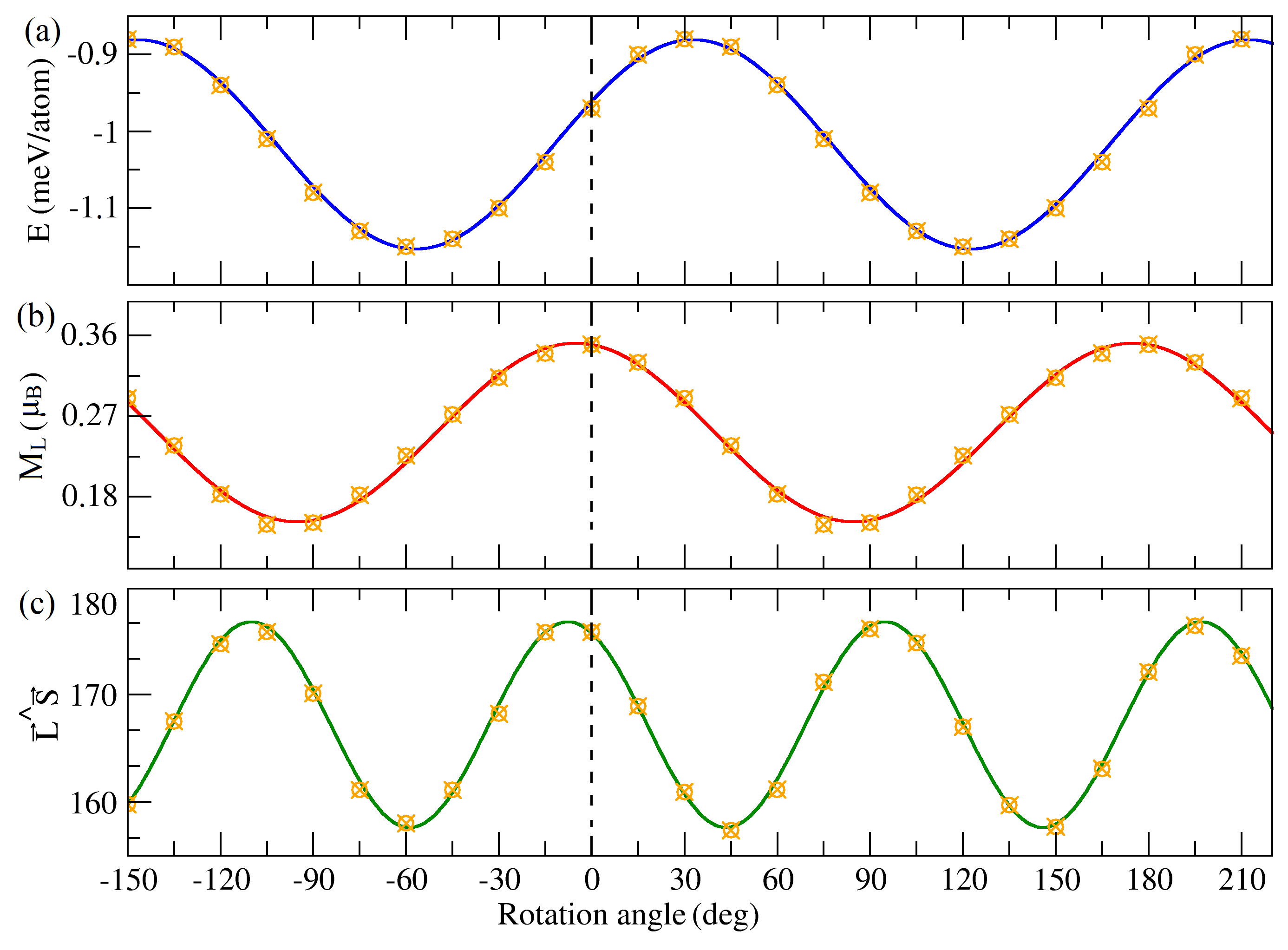}
\caption{  a) Single-ion anisotropy energy depending on the spin direction in the $xz$-plane. The experimental spin direction $\boldsymbol{\gamma} \sim (0.57, 0, 0.82)$ is chosen as zero. b) Angular dependence of the orbital moment. (c) The angle between the spin and orbital moments. }
\label{fig:SIA}
\end{figure}
In BaMoP$_2$O$_8$, magnetic anisotropy comprises the on-site term represented by the single-ion tensor $A^{\mu \nu}_{i}$, and the intersite term, represented by the traceless $\Gamma^{\mu \nu}_{ij}$ tensor. We begin with the first part. In order to evaluate the on-site anisotropic term $A^{\mu \nu}_{i}$, we rotate the magnetic moment of an individual Mo atom within the $xz$-plane, while keeping the magnetic moments of other Mo atoms along the $y$ direction, which cancels all isotropic exchange interactions. The resulting energies given with respect to the experimental spin direction $\boldsymbol{\gamma}$ are shown in Fig.~\ref{fig:SIA}. This direction is not the energy minimum. Instead, we find $\mathbf{e}_{\min}=(0.98,0,-0.18)$ as the easy-axis direction according to the on-site anisotropy, and $\Delta E_A = E_{\max}-E_{\min}= 0.28$\,meV. 

As for the intersite anisotropy, its individual components are determined using the the mapping procedure~\cite{xiang2011} similarly to the isotropic exchange interactions $J_{ij}$, but in this case we use non-collinear spin configurations and take the spin-orbit coupling effects into account. For the dominant coupling $J_2$, we find (in meV):

\[\Gamma^{\mu \nu}_{2}= \left( \begin{array}{rrr} 
-0.033 & 0.000 & -0.090  \\
 0.000 & 0.036 &  0.000  \\
-0.090 & 0.000 & -0.003   
\end{array} \right). 
\]
This anisotropy tensor is compatible with the symmetry of the Mo--Mo bond, the two-fold rotation axis along $b$ that cancels out all off-diagonal components but $xz$ and $zx$.

By combining this anisotropy term with the on-site anisotropy, we find only a small change in the easy direction, which still does not fit to the experimental one. 
On the other hand, the experimental spin direction clearly correlates with the direction, where the maximum value of the orbital moment is achieved (Fig.~\ref{fig:SIA}b). This is in agreement with the conventional argument that magnetocrystalline anisotropy is fully determined by the orbital moment~\cite{bloch1931,bruno1989}. Moreover, $\boldsymbol{\gamma}$ is close to the special direction, where spin and orbital moments are collinear, i.e., it is a principal direction of the $g$-tensor. One can, therefore, elucidate the easy direction of BaMoP$_2$O$_8$ in terms of the $g$-tensor anisotropy. 

\section{Discussion and Summary}
\label{sec:discussion}
BaMoP$_2$O$_8$ is a rare material showing a triangular arrangement of the $4d$ ions. However, monoclinic distortion of the crystal structure has a strong effect on its magnetic interactions and renders the spin lattice quasi-one-dimensional, with $S=1$ chains running along the $b$ direction. Experimentally, BaMoP$_2$O$_8$ reveals two features typical for quantum magnets, namely: i) the broad susceptibility maximum due to the short-range order around 46\,K that precedes the long-range magnetic order formed below $T_N\simeq 21$\,K; and ii) the reduction in the ordered moment due to quantum fluctuations. Both features are likely a combined effect of the magnetic one-dimensionality driven by the monoclinic distortion, and frustration caused by the competing interchain couplings $J_1$.

Stripe magnetic order observed in BaMoP$_2$O$_8$ is not unexpected in spatially anisotropic triangular antiferromagnets approaching the one-dimensional limit~\cite{starykh2007}. The ordered state features antiparallel spins along the spin chains formed by the leading coupling $J_2$. Such chains are decoupled on the mean-field level, because the interchain couplings $J_1$ are fully frustrated. Nevertheless, collinear interchain order forms, with ferromagnetic spin alignment on half of the $J_1$ bonds. In the case of spin-$\frac12$, this stripe order can be stabilized by quantum fluctuations~\cite{pardini2008,bishop2009,ghamari2011}, although a disordered phase with stripe fluctuations was also reported in the $J_1\ll J_2$ limit~\cite{heidarian2009,reuther2011,schmidt2014,ghorbani2016}. 

The $S=1$ magnets should be less prone to the stripe order in the $J_1\ll J_2$ limit, because quantum fluctuations are reduced~\cite{li2012}. Moreover, decoupled chains form the Haldane phase protected by a spin gap, and a sizable $J_1/J_2\geq 0.3=0.4$~\cite{pardini2008,gonzalez2017} would be needed to induce the ordering. Since BaMoP$_2$O$_8$ with its $J_1/J_2\simeq 0.1$ is clearly below this threshold value, we conclude that $J_1$ can't cause long-range order in this system. 

The interlayer coupling $J_3$ is a more plausible candidate, because it is non-frustrated and couples the spin chains more efficiently. Our estimate of $J_3/J_2\simeq 0.043$ is surprisingly close to the reported threshold value of $J_{\rm inter}/J_{\rm intra}=0.042-0.044$~\cite{kim2000,matsumoto2001,moukouri2011,gonzalez2017}, although one should keep in mind that the interaction $J_3$ acts to induce the long-range order in the $bc$ plane only, whereas the couplings in the $ab$ plane remain frustrated. Finally, the single-ion anisotropy of $A/J_2\simeq 0.06$ is too weak to close the Haldane gap on its own, as higher values of $A/J_2\geq 0.31$ would be required in the absence of interchain couplings~\cite{albuquerque2009}. Therefore, we conclude that the combined effect of easy-axis anisotropy and $J_3$~\cite{wierschem2014} is required to stabilize long-range magnetic order in BaMoP$_2$O$_8$. The remarkably small size of the ordered domains, on the order of 10 interatomic distances, corroborates the frustrated nature of the system and is in agreement with recent predictions of the reduced correlation length at $J_1\ll J_2$~\cite{weichselbaum2011,reuther2011}.

BaMoP$_2$O$_8$ shows close similarity to other monoclinically distorted yavapaiites. For example, KTi(SO$_4)_2$ is a quasi-one-dimensional spin-$\frac12$ antiferromagnet~\cite{nilsen2015}. In both Ti and Mo compounds, orbital order on the transition-metal site supports magnetic one-dimensionality. The anisotropic nature of the magnetic orbital(s) leads to a large difference between $J_1$ and $J_2$ in the monoclinic structure, because the superexchange pathway of $J_2$ is more linear and constrained by the two-fold symmetry axis, whereas the $J_1$ pathway has no symmetry constraints and thus departs from linearity. On the other hand, $4d$ ions with more isotropic orbitals, such as Rh$^{4+}$ with its putative $j_{\rm eff}=\frac12$ state~\cite{calder2015}, may produce a more regular triangular spin lattice even in the presence of the monoclinic distortion. Given the similar ionic radii of Mo$^{4+}$ and Rh$^{4+}$, accommodating Rh in the yavapaiite structure is an interesting direction for further studies. It may also be possible to suppress the monoclinic distortion itself. For example, among the Fe-based yavapaiites only KFe(SO$_4)_2$ is monoclinic, whereas RbFe(SO$_4)_2$ and CsFe(SO$_4)_2$ show robust trigonal symmetry that imposes $J_1=J_2$~\cite{serrano1998}.

The local physics of Mo$^{4+}$ may be of interest, too. The $4d^2$ electronic configuration leads to an orbitally degenerate scenario for an octahedrally coordinated ion. This orbital degeneracy is usually lifted by the formation of Mo--Mo bonds, as in the dimerized Y$_2$Mo$_2$O$_7$~\cite{thygesen2017} or trimerized Zn$_2$Mo$_3$O$_8$~\cite{mccarroll1957}. However, large Mo--Mo distances and relatively weak electronic interactions in BaMoP$_2$O$_8$ prevent the dimerization, such that local distortions of the MoO$_6$ octahedron are left to choose the orbital configuration. The orbital moment is far from being fully quenched and amounts to about 0.35\,$\mu_B$. Despite this sizable orbital moment, magnetic interactions between the Mo$^{4+}$ ions are nearly isotropic. Only a weak exchange anisotropy is revealed by our \textit{ab initio} studies. An interesting question at this point is whether such an orbital state is generic for all Mo$^{4+}$ phosphates, or different local distortions of MoO$_6$ may cause a variable orbital ground state and stronger exchange anisotropy.

In summary, we used thermodynamic measurements, neutron diffraction, and \textit{ab initio} calculations to explore the magnetic behavior of yavapaiite-structured BaMoP$_2$O$_8$ with the triangular arrangement of the $4d^2$ Mo$^{4+}$ ions. The monoclinic distortion has a drastic influence on the magnetic scenario and breaks down the triangular spin lattice into weakly coupled $S=1$ chains. The stripe order with the propagation vector $\kv=(\frac12,\frac12,\frac12)$ forms below $T_N\simeq 21$\,K following the combined effect of magnetic one-dimensionality and easy-axis anisotropy. The ordered moment of 1.42(9)\,$\mu_B$ at 1.5\,K shows significant reduction compared to the spin-only value due to the interplay of quantum fluctuations and spin-orbit coupling.

\acknowledgments
We acknowledge fruitful discussions with Philipp Gegenwart and Anton Jesche, and the provision of neutron beamtime by the ILL. We also thank the staff of ID22 at the ESRF for their continuous technical support during the synchrotron measurements. The work in Augsburg was supported by the Federal Ministry for Education and Research through the Sofja Kovalevskaya Award of Alexander von Humboldt Foundation and by the German Science Foundation via TRR80. The work of D.I.B. was funded by RFBR according to the research project No 18-32-00018. The work or V.V.M. was supported by the Russian Science Foundation, Grant No 18-12-00185.

\end{document}